\documentclass[aps,prc,twocolumn,showpacs,preprintnumbers,amsmath,amssymb,
epsf,superscriptaddress,groupedaddress,nofootinbib,tightenlines]{revtex4}

\usepackage{graphicx}
\usepackage{amsmath}
\usepackage{bm}

\def\beqn{\begin{eqnarray}}
\def\eeqn{\end{eqnarray}}
\def\barr{\begin{array}}
\def\earr{\end{array}}
\def\btab{\begin{tabular}}
\def\etab{\end{tabular}}
\def\bite{\begin{itemize}}
\def\eite{\end{itemize}}
\def\bcen{\begin{center}}
\def\ecen{\end{center}}

\def\eq{\begin{equation}}
\def\ee{\end{equation}}
\def\nn{\nonumber}

\def\kdagger{K\hspace{-0.3cm}/}
\def\ndagger{n\hspace{-0.2cm}/}

\def\keldagger{k\hspace{-0.2cm}/}

\def\q2dagger{q_2\hspace{-0.35cm}/\;}

\begin{document}


\title{Beam normal spin asymmetry in the quasi-RCS approximation}
\author{M. Gorchtein}
\affiliation{Genoa University, Department of Physics, 16146 Genoa, Italy}
\affiliation{California Institute of Technology, Pasadena, CA 91125, USA}
\email{gorshtey@caltech.edu}
\begin{abstract}
The two-photon exchange contribution to the single spin asymmetries 
with the spin orientation normal to the reaction plane is discussed for 
elastic electron-proton scattering in the equivalent photon approximation. 
In this case, hadronic part of the two-photon exchange amplitude describes 
real Compton scattering (RCS). We show that in the case of 
the beam normal spin asymmetry, this approximation selects only the photon 
helicity flip amplitudes of RCS. At low energies, we make use of unitarity 
and estimate the contribution of the $\pi N$ multipoles to the photon 
helicity flip amplitudes. In the Regge regime, QRCS approximation
allows for a contribution from two pion exchange, and we provide an estimate 
of such contributions. We furthermore discuss the possibility of the 
quark and gluon GPD's contributions in the QRCS kinematics. 
\end{abstract}
\maketitle
\section{Introduction}
\label{sec:intro}
Recently, the new polarization transfer data for the electromagnetic form 
factors ratio $G_E/G_M$ \cite{gegm_exp} raised a lot of interest for the 
two photon exchange (TPE) physics in elastic electron proton scattering. These 
new data appeared to be incompatible with the Rosenbluth data 
\cite{rosenbluth}. A possible way to reconcile the two data sets was 
proposed \cite{marcguichon}, which consists in a more precise account on the 
TPE amplitude, the real part of which enters the radiative 
corrections to the cross section (Rosenbluth) and the polarization cross 
section ratio in a different manner. At present, only the IR divergent part 
of the two photon exchange contribution, 
corresponding to one of the exchanged photon beeing soft,
is taken into the experimental analysis \cite{motsai_maximon}. 
Two model calculations exist for the real part of the 
TPE amplitude \cite{melnichuk,suki}, and they qualitatively confirm that 
the exchange of two hard photons may be responsible for this discrepancy. 
In order to extract the electric form factor in the model independent 
way, one has thus to study the general case of Compton scattering with two 
spacelike photons. These two photon contributions are important for the 
electroweak sector, as well.\\
\indent
In view of this interest, parity-conserving
single spin asymmetries 
in elastic $ep$-scattering with the spin orientation normal to 
the reaction plane regain an attention. These observables are directly 
related to the imaginary part of the TPE amplitude and have been an object 
of theoretical studies in the 1960's and 70's \cite{derujula}. 
By analyticity, the real part 
of the TPE amplitude is given by a dispersion integral over its imaginary 
part. Therefore, a good understanding of this class of observables is 
absolutely necessary. Recently, first measurements of the beam normal 
spin asymmetry $B_n$\footnote{In the literature, also the $A_n$ notation 
for beam normal spin asymmetry or vector analyzing power was adopted.} 
have been performed in different kinematics \cite{bn_exp}. \\
\indent
Though small (several to tens ppm), this asymmetry can be measured with a 
precision of fractions of ppm. 
Before implementing different models for the real part, where an additional 
uncertainty comes from the dispersion integral over the imaginary part, one 
should check the level of understanding of the imaginary part of TPE. These 
checks have been done for the existing data. Inclusion of the elastic 
(nucleon) intermediate state only \cite{afanas} 
led to negative asymmetry of several ppm in 
the kinematics of SAMPLE experiment but was not enough to describe the data. 
The description of the beam 
normal spin asymmetry within a phenomenological model which uses the full set 
of the single pion electroproduction \cite{marcbarbara} did not give 
satisfactory description at any of the available kinematics. Especially 
intriguing appears the situation with the SAMPLE data with electron lab energy 
$E_{lab}=200 $ MeV, which is just above the 
pion production threshold where the theoretical input is well understood. 
On the other hand, an EFT calculation without dynamical pions 
\cite{diaconescu} was somewhat surprisingly very successful in describing this 
kinematics for $B_n$. This success suggests that to the given order of 
chiral perturbation theory, the role of 
the dynamical pions for this observable might be not too large. 
Finally, a recent calculation within the leading logarithm approximation 
appeared, where only few dominant multipoles were used as input 
\cite{kobushkin}. Surprizingly, the authors  of \cite{kobushkin} were able to 
describe all the experimental data at very different beam energies and 
scattering angles, while the full calculation of \cite{marcbarbara} failed to 
describe any of them. \\
\indent
Though even at low energies the situation with the imaginary part of TPE 
amplitude is by far not clear, an attention has to be paid to high energies, 
as well, since the dispersion integral which would give us its real part, 
should be performed over the full energy range. Due to relative ease of 
measuring $B_n$ within the framework of parity violating electron scattering, 
new data from running and upcoming experiments \cite{bn_proposals} will 
stimulate further theoretical investigations of this new observable. A 
calculation of $B_n$ in hard kinematics regime at high energy and momentum 
transfer was performed recently in the framework of generalized parton 
distributions (GPD's) and resulted in asymmetries of $\sim1.5$ ppm. 
\cite{jamarcguichon}.\\ 
\indent
Since a ppm effect measurement at high momentum transfers is an extremely 
complicated task, the forward kinematics seems more favorable. 
For this kinematics, a calculation exists \cite{afanas_qrcs}, where an 
observation 
is made that the contribution of the situation where the exchanged 
photons are nearly real and overtake the external electron kinematics is 
enhanced as $\ln^2(-t/m^2)\sim100$, with $m$ the electron mass and $t<0$ the 
elastic momentum transfer. Making use of the optical theorem, the authors 
obtained estimates of $B_n$ in this kinematics as large as 25-35 ppm. 
The calculation of \cite{ja_qrcs} appeared afterwards showed that this 
result is not adequate, since the squared log term can only contribute with 
the Compton scattering amplitudes with the photon helicity flip. 
Indeed, this conclusion was confirmed in Ref. \cite{afanas_erratum}, 
the corrected version of \cite{afanas_qrcs}.
It was noticed that the $\ln^2(-t/m^2)$ term should vanish in the forward 
kinematics due to gauge invariance, 
and the leading term is governed by a single (though still large) 
log term, $\ln(-t/m^2)\sim10$. The predictions for $B_n$ shifted 
correspondingly to $\approx-(4-6)$ ppm and agree with the preliminary data 
from HAPPEX experiment \cite{bn_proposals}.\\
\indent
In order to provide an estimate of $B_n$, we use 
the equivalent photon or quasi-real Compton scattering approximation which is 
caused by the hard collinear kinematics, responsible for this $\ln^2(-t/m^2)$ 
enhancement. In this approximation, the leading contribution comes from the 
kinematical region with both exchanged photons are almost real. 
The hadronic tensor is taken at this kinematical point 
and can be taken out of the integration. For the hadronic tensor, 
we adopt the most general real Compton scattering amplitude and demostrate that 
the contribution of the cross section (i.e., photon helicity conserving 
amplitudes) vanishes in the QRCS approximation. The remaining contributions are 
related to the photon helicity flip amplitudes. We provide the calculation 
of $B_n$ at low energies, where $\pi N$ intermediate states are expected to 
be the dominant contributions. In this kinematics, the asymmetry can be related 
to the pion photoproduction multipoles, and we discuss the relative 
contributions of different multipoles. At higher energies, we discuss forward 
kinematics, that is Regge regime, and note that the combinations of the 
RCS amplitudes appearing in the expression of $B_n$ are related to $2\pi$ 
exchange in the $t$-channel. We provide an estimate of such a contribution but 
conclude that these are negligibly small. We furthermore discuss the hard 
regime and discuss possible contributions which might be enhanced by the large 
logarithms originating from the QRCS kinematics.
\\
\indent
The article is organized as follows: in Section \ref{sec:el_ampl}, 
we define the kinematics, general $ep$-scattering amplitude and the 
observables of interest; in Section \ref{sec:tpe}, the two photon exchange 
mechanism and the photons kinematics is studied; the equivalent photons or 
quasi real Compton scattering approximation and its implementation for the 
case of $B_n$ is given in Section \ref{sec:qrcs}; we present our results in 
Section \ref{sec:results} and conclude with a short summary. 
\section{Elastic $ep$-scattering amplitude}
\label{sec:el_ampl}
In this work, we consider elastic electron-proton scattering process 
$e(k)+p(p)\to e(k')+p(p')$ for which we define:
\beqn
P&=&\frac{p+p'}{2}\nn\\
K&=&\frac{k+k'}{2}\nn\\
q&=&k-k'\;=\;p'-p,
\eeqn
and choose the invariants $t=q^2<0$\footnote{In elastic $ep$-scattering, the 
usual notation for the momentum transfer is $Q^2=-q^2$ but we prefer the 
more general notation $t$ to avoid confusion with the incoming and outgoing 
photon virtualities in forward doubly virtual Compton scattering we will 
be concerned in the following.} and $\nu=(P\cdot K)/M$ as the 
independent variables. $M$ denotes the nucleon mass. They are related to the 
Mandelstam variables $s=(p+k)^2$ and $u=(p-k')^2$ through $s-u=4M\nu$ and 
$s+u+t=2M^2$. For convenience, we also introduce the usual polarization 
parameter $\varepsilon$ of the virtual photon, which can be related to the 
invariants $\nu$ and $t$ (beglecting the electron mass $m$):
\beqn
\varepsilon\,=\,\frac{\nu^2-M^2\tau(1+\tau)}{\nu^2+M^2\tau(1+\tau)},
\eeqn
with $\tau=-t/(4M^2)$. Elastic scattering of two spin $1/2$ particles is 
described by six independent amplitudes. Three of them do not flip the 
electron helicity \cite{marcguichon},
\beqn
T_{no\;flip}&=&\frac{e^2}{-t}
\bar{u}(k')\gamma_\mu u(k)\label{f1-3}
\\
&\cdot&
\bar{u}(p')
\left(\tilde{G}_M \gamma^\mu\,-\,
\tilde{F}_2\frac{P^\mu}{M}\,+\,
\tilde{F}_3\frac{\kdagger P^\mu}{M^2}\right)u(p),\nn
\eeqn
while the other three are electron helicity flipping and thus have in 
general the order of the electron mass $m$ \cite{jamarcguichon}:
\beqn
T_{flip}&=&\frac{m}{M}\frac{e^2}{-t}
\left[
\bar{u}(k')u(k)\cdot\bar{u}(p')\left(\tilde{F}_4\,+\,
\tilde{F}_5\frac{\kdagger}{M}\right)u(p)\right.\nn\\
&&\;\;\;\;\;\;\;\;\;\;\;+\;
\tilde{F}_6\bar{u}(k')\gamma_5u(k)\cdot\bar{u}(p')\gamma_5u(p)
\Big]
\label{f4-6}
\eeqn
\indent
In the one-photon exchange (Born) approximation, two of the six amplitudes 
match with the electromagnetic form factors,
\beqn
\tilde{G}_M^{Born}(\nu,t)&=&G_M(t),\nn\\
\tilde{F}_2^{Born}(\nu,t)&=&F_2(t),\nn\\
\tilde{F}_{3,4,5,6}^{Born}(\nu,t)&=&0
\eeqn
where $G_M(t)$ and $F_2(t)$ are the magnetic and Pauli form factors, 
respectively. For further convenience we define also 
$\tilde{G}_E\,=\,\tilde{G}_M-(1+\tau)\tilde{F}_2$ and 
$\tilde{F}_1\,=\,\tilde{G}_M-\tilde{F}_2$ which in the Born approximation 
reduce to electric form factor $G_E$ and Dirac form factor $F_1$, 
respectively. For a beam polarized normal to the scattering plane, one can 
define a single spin asymmetry,
\beqn
B_n\,=\,
\frac{\sigma_\uparrow-\sigma_\downarrow}{\sigma_\uparrow+\sigma_\downarrow},
\eeqn
where $\sigma_\uparrow$ ($\sigma_\downarrow$) denotes the cross sesction for 
an unpolarized target and for an electron beam spin parallel (antiparallel) 
to the normal polarization vector defined as 
\beqn
S_n^\mu\,=\,
\left(0,\frac{[\vec{k}\times\vec{k'}]}{|\vec{k}\times\vec{k'}|}\right),
\eeqn
normalized to $(S\cdot S)=-1$. Similarly, one defines the target normal spin 
asymmetry $A_n$. It has been shown in the early 1970's \cite{derujula} that 
such asymmetries are directly related to the imaginary part of the $T$-matrix.
Since the electromagnetic form factors and the one-photon exchange amplitude
are purely real, $B_n$ obtains its finite contribution to leading order in the 
electromagnetic constant $\alpha_{em}$ from an interference between the Born 
amplitude and the imaginary part of the two-photon exchange amplitude.
In terms of the amplitudes of Eqs.(\ref{f1-3},\ref{f4-6}), the beam normal 
spin asymmetry is given by:
\beqn
B_n&=&-\frac{m}{M}\sqrt{2\varepsilon(1-\varepsilon)}\sqrt{1+\tau}
\left(\tau G_M^2+\varepsilon G_E^2\right)^{-1}\nn\\
&\cdot&
\left\{
\tau G_M {\rm Im}\tilde{F}_3\,+\,G_E {\rm Im}\tilde{F}_4
\,+\,F_1\frac{\nu}{M}{\rm Im}\tilde{F}_5
\right\}.
\label{eq:bn_general}
\eeqn
\indent
For completeness, we also give here the expression of target normal spin 
asymmetry $T_n$\footnote{Also $A_n$ notation for target normal spin asymmetry 
exists in the literature.} in terms of invariant amplitudes:
\beqn
T_n&=&\sqrt{2\varepsilon(1+\varepsilon)}\sqrt{\tau}
\left(\tau G_M^2+\varepsilon G_E^2\right)^{-1}\label{eq:an_general}
\\
&\cdot&
\left\{
(1+\tau)\left[F_1{\rm Im}\tilde{F}_2-F_2{\rm Im}\tilde{F}_1\right]\right.\nn\\
&&\left.\,+\,\left(\frac{2\varepsilon}{1+\varepsilon}G_E-G_M\right)
\frac{\nu}{M}{\rm Im}\tilde{F}_3
\right\}.\nn
\eeqn
\section{Two photon exchange}
\label{sec:tpe}
\begin{figure}[h]
{\includegraphics[height=2cm]{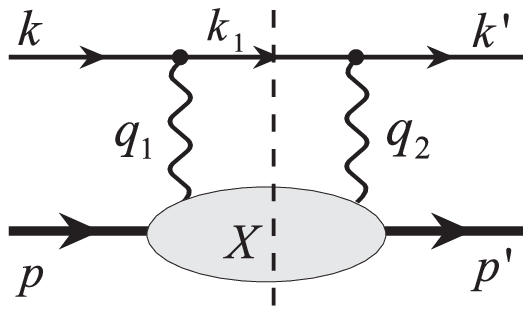}}
\caption{Two-photon exchange diagram.}
\label{boxgraph}
\end{figure}
\begin{figure}[h]
{\includegraphics[height=7cm]{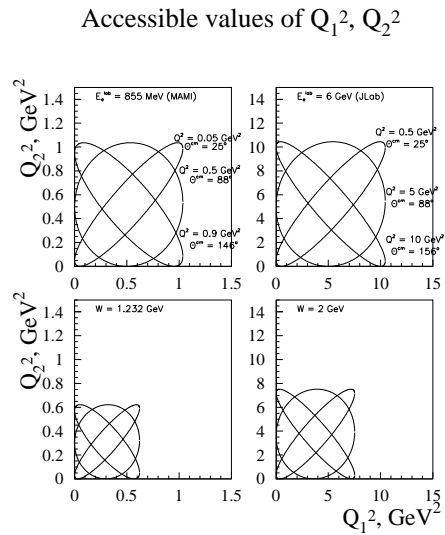}}
\caption{Kinematically allowed values of the photon vitualities $Q_{1,2}^2$.}
\label{fig:kinbounds}
\end{figure}

The imaginary part of the two-photon exchange (TPE) graph in Fig.\ref{boxgraph}
is given by 
\beqn
{\rm Im} {\cal{M}}_{2\gamma}&=&e^2
\int\frac{|\vec{k}_1|^2d|\vec{k}_1|d\Omega_{k_1}}{2E_1(2\pi)^3}
\bar{u}'\gamma_\nu(\keldagger_1+m)\gamma_\mu u\nn\\
&&\;\;\;\;\cdot\,\frac{1}{Q_1^2Q_2^2}W^{\mu\nu}(w^2,Q_1^2,Q_2^2),
\label{eq:im2gamma}
\eeqn
where $W^{\mu\nu}(w^2,Q_1^2,Q_2^2)$ is the imaginary part of doubly virtual 
Compton scattering tensor. $Q_1^2$ and $Q_2^2$ denote the virtualities of 
the exchanged photons in the TPE diagram, and $w$ is the invariant mass of 
the intermediate hadronic system. We next study the kinematics of the 
exchanged photons. Neglecting the small electron mass and using the c.m. frame 
of the electron and proton, one has:
\beqn
Q_{1,2}^2\,=\,2|\vec{k}||\vec{k}_1|(1-\cos\Theta_{1,2}),
\eeqn
with $|\vec{k}|={s-M^2\over2\sqrt{s}}\equiv k$ the three momentum of the 
incoming (and outgoing) eletron, 
$|\vec{k}_1|=\sqrt{({s-w^2+m^2\over2\sqrt{s}})^2-m^2}$ that of the 
intermediate electron, and 
$\cos\Theta_2=\cos\Theta\cos\Theta_1+\sin\Theta\sin\Theta_1\cos\phi$. 
The kinematically allowed values of the virtualities of the exchanged photons 
(the restriction is due to the fact that the intermediate electron is on-shell)
are represented by the internal area of the ellypses shown in 
Fig. \ref{fig:kinbounds}. 
The ellypses are drawn inside a square whose side is 
defined through the external kinematics ($k$) and the invariant mass of the 
intermediate hadronic state ($w^2$ or $k_1$), while the form solely by the 
scattering angle. If choosing higher values of the mass of the hadronic 
system $w^2<s$, it leads to scaling the size of the ellypse by a factor of 
$s-w^2\over s-M^2$. In the limit $w^2=(\sqrt{s}-m_e)^2$, the ellypses shrink 
to a point at the origin and both photons are nearly real. This is not a 
soft photon (IR) singularity, however, since the real photons' energy 
remains large 
enough in order to provide the transition from nucleon with the mass $M$ to 
the intermediate state $X$ with the mass $w$. Instead, the intermediate 
electron is soft, $k_1^\mu\approx(m_e,\vec{0})$, therefore this kind of 
kinematics does not lead to an IR divergency which can only occur if the 
intermediate hadronic state is the nucleon itself. In the following we are 
going to study this kinematical situation in more detail.
\section{Quasi-RCS approximation}
\label{sec:qrcs}
We consider the kinematical factors under the integral over the electron phase 
space of Eq.(\ref{eq:im2gamma}) at the upper limit of the integration over the 
invariant mass of the intermediate hadronic state, $w\to\sqrt{s}-m_e$:
\beqn
\frac{\vec{k}_1^2}{E_1}
\frac{1}{Q_1^2Q_2^2}\,\sim\,
\frac{1}{4k^2E_1}
\,\sim\,{1\over m} {1\over4k^2},
\eeqn
In this range of 
the hadronic mass $w$, the exchanged photons are real, and the contribution of 
real Compton scattering (RCS) to the $2\gamma$-exchange graph is enhanced by 
a factor of $1/m$ (it is not a singularity since $B_n$ has a factor of $m$ 
in front). \\
\indent
We next rewrite the contraction of the hadronic and the leptonic tensors in 
Eq.(\ref{eq:im2gamma}) identically as 
\beqn
&&l_{\mu\nu}W^{\mu\nu}(W^2,Q_1^2,Q_2^2)\,=\,l_{\mu\nu}^{(0)}W^{\mu\nu}(s,0,0)
\nn\\
&&+l_{\mu\nu}^{(1)}W^{\mu\nu}(s,0,0)\nn\\
&&+l_{\mu\nu}
\left[W^{\mu\nu}(W^2,Q_1^2,Q_2^2)-W^{\mu\nu}(s,0,0)\right],
\label{eq:rcstensor_reg}
\eeqn
where we expanded the leptonic tensor $l_{\mu\nu}=l_{\mu\nu}^0+l_{\mu\nu}^1$ 
with
\beqn
l_{\mu\nu}^0&=&m\bar{u}'\gamma_\nu\gamma_\mu u\nn\\
l_{\mu\nu}^1&=&\bar{u}'\gamma_\nu\keldagger_1\gamma_\mu u.
\eeqn

In Eq.(\ref{eq:rcstensor_reg}), only the first term does not vanish in the 
QRCS limit, while the second and third are at least linear in $k_1$.
In the limit $W^2\to s$, one has $\keldagger_1\to m\gamma_0$, which is to be 
compared to the term $m\cdot 1$ in $l_{\mu\nu}^0$. Though on the first 
glance on the first glance they are of the same order, the additional 
$\gamma$-matrix picks an extra electron mass $m$ from one of the projectors 
$\keldagger+m$ when performing the summation over electron spins. 
Quasi-real Compton scattering (QRCS) approximation 
consists in assuming the first term to be dominant due to the kinematical 
enhancement under the integral and in neglecting the second one. 
In general, this kind of contributions coming from QRCS kinematics 
will allways be present in the full calculation, since the second term in 
Eq.(\ref{eq:rcstensor_reg}) is constructed in such a way that the resulting 
integral is regular at the QRCS point. In the following, the QRCS 
approximation will be used.
Hence, the hadronic tensor can be taken out of the integration over 
the electron phase space. The remaining integral is
\beqn
I_0&=&\int\frac{d^3\vec{k}_1}{2E_1(2\pi)^3}\frac{1}{Q_1^2Q_2^2},
\eeqn
\indent
The result of the integration reads:
\beqn
I_0\;=\;\frac{1}{-32\pi^2t}
\left\{
\ln^2\left(\frac{-t}{m^2}\frac{E_{thr}^2}{E^2}\right)\,+\,
8Sp\left({{E_{thr}}\over E}\right)\right\},
\eeqn
where $Sp(x)$ is the Spence or dilog function, 
$Sp(x)=-\int_0^1{dt\over t}\ln(1-xt)$ with $Sp(1)=\pi^2/6$. In the high 
energy limit, ${{E_{thr}}\over E}\to1$, we recover the result of 
Ref.\cite{afanas_qrcs}. For the details, we address the reader to the Appendix.
This $\ln^2\left(\frac{-t}{m^2}\right)$ factor serves as the justification of 
the QRCS approximation. The full result may be represented as a ``Taylor 
expansion'' in powers of $\ln\left(\frac{-t}{m^2}\right)\sim10$. By studying 
the QRCS approximation we show in a model independent way that the leading term 
in this expansion is quadratic. Since this leading term is large, we can (under 
ceretain kinematical conditions) neglect further terms. \\
\indent
The RCS tensor may be taken for instance in the basis of Prange \cite{prange} 
or, equivalently of Berg and Lindner \cite{berglindner},
\beqn
{W}^{\mu\nu}_{RCS}&=&\bar{N}'
\left\{
\frac{P'^\mu P'^\nu}{P'^2}(B_1\,+\,\kdagger B_2)\;+\;
\frac{n^\mu n^\nu}{n^2}(B_3\,+\,\kdagger B_4)\right.\nn\\
&&\;\;\;\;\;\;
\;+\frac{P'^\mu n^\nu\,-\,n^\mu P'^\nu}{P'^2n^2}\,i\gamma_5 B_7\nn\\
&&\;\;\;\;\;\;
\left.\;+\frac{P'^\mu n^\nu\,+\,n^\mu P'^\nu}{P'^2n^2}\,\ndagger B_6
\right\}\,N,
\label{eq:rcstensor}
\eeqn
with the vectors defined as $P'=P-{P\cdot K\over K^2}K$, 
$n^\mu=\varepsilon^{\mu\nu\alpha\beta}P_\nu K_\alpha q_\beta$ such that 
$P'\cdot K = P'\cdot n= n\cdot K=0$. The amplitudes $B_i$ are functions of 
$\nu$ and $t$. This form of Compton tensor is 
convenient due to the simple form of the tensors appearing in 
Eq. (\ref{eq:rcstensor}). \\
\indent
Before contracting the leptonic and hadronic tensors,
we notice that the amplitude $B_7$ can only contribute to the invariant 
amplitude $\tilde{F}_6$, since it contains the $\bar{N}'\gamma_5N$ structure.
$\tilde{F}_6$ does not contribute at leading order in $m$
to neither obserevable of interest, therefore $B_7$ will
be neglected in the following. The remaining tensors in Eq.(\ref{eq:rcstensor})
 are symmetric in indices $\mu\nu$.
\beqn
{\rm Im} {\cal{M}}_{2\gamma}^{QRCS}&=&e^2
mI_0\bar{u}'u
W_{RCS}^{\mu\nu}g_{\mu\nu}.
\label{eq:tensor_contraction}
\eeqn
\indent
Finally, we identify 
different terms in Eq. (\ref{eq:tensor_contraction}) 
with the structures, together with amplitudes $\tilde{F}_{1,\dots6}$
parametrizing elastic $ep$-scattering amplitude in Eqs. (\ref{f1-3},\ref{f4-6})
and find for the invariant amplitudes for 
the elastic electron-proton scattering in the QRCS approximation:
\beqn
{\rm Im} \tilde{G}_M^{QRCS}&=&
{\rm Im} \tilde{F}_2^{QRCS}\;=\;
{\rm Im} \tilde{F}_3^{QRCS}\;=\;0
\label{f3}\\
{\rm Im} \tilde{F}_4^{QRCS}&=&-MtI_0{\rm Im}(B_1+B_3)\label{f4}\\
{\rm Im} \tilde{F}_5^{QRCS}&=&-M^2tI_0{\rm Im}(B_2+B_4)\label{f5}
\eeqn
\indent

We obtain for $B_n$ in the QRCS approximation:
\beqn
B_n&=&mtI_0\sqrt{2\varepsilon(1-\varepsilon)}\sqrt{1+\tau}
\left(\tau G_M^2+\varepsilon G_E^2\right)^{-1}\nn\\
&\cdot&
\left\{
G_E {\rm Im}(B_1+B_3)
\,+\,\nu F_1{\rm Im}(B_2+B_4)
\right\}.
\label{eq:B_n_Bi}
\eeqn

The result of Eq.(\ref{eq:B_n_Bi}) is obtained by only using the assumption 
that the QRCS kinematics dominate the integral in Eq.(\ref{eq:im2gamma}).
The combinations of the RCS amplitudes appearing in the final result can be 
furthermore expressed in terms of the helicity amplitudes of real Compton 
scattering. With these latter defined as 
$T_{\lambda'_\gamma\lambda'_N,\lambda_\gamma\lambda_N,}\,\equiv\,\varepsilon'^{*\nu}_{\lambda'_\gamma}\varepsilon^\mu_{\lambda'_\gamma}W_{\mu\nu}^{RCS}$, 
one has \cite{javcs,lvov}:
\beqn
B_1+B_3
&=&-\frac{1}{\sqrt{-t}}
\left(T_{-1-{1\over2},1{1\over2}}+T_{1-{1\over2},-1{1\over2}}\right)\nn\\
&-&\frac{2M}{M^4-su}T_{1{1\over2},-1{1\over2}}
\label{eq:b1pb3_helampl}\\
B_2+B_4
&=&\frac{2M}{s-M^2}\frac{1}{\sqrt{-t}}
\left(T_{-1-{1\over2},1{1\over2}}+T_{1-{1\over2},-1{1\over2}}\right)\nn\\
&+&\frac{2}{M^4-su}\frac{s+M^2}{s-M^2}T_{1{1\over2},-1{1\over2}}
\label{eq:b2pb4_helampl}
\eeqn
\indent
As can be seen, only photon helicity-flipping amplitudes enter the final result 
for $B_n$ in the QRCS approximation. This is the main result of this work. 

\section{Results}
\label{sec:results}
In this section, we consider the impact of the QRCS contributions onto the 
beam normal spin asymmetry in different kinematics: low energies ($\pi N$ 
intermediate states), high energies and forward angles, i.e. Regge regime 
(2$\pi$ exchange in the $t$-channel) and hard regime (handbag diagrams and 
two gluon exchange).
\subsection{Low energies: $\pi N$ multipoles}
Above the pion production threshold, the imaginary part of the RCS helicity 
amplitudes can be related to the pion photo- and electro-production multipoles. 
These relations read \cite{lvov, javcs} for the three amplitudes entering $B_n$:
\beqn
{\rm Im}T_{-1-{1\over2},1{1\over2}}&=&\sin{\Theta\over2}16\pi q_\pi\sqrt{s}
\sum_{k\geq0}(k+1)^2\nn\\
&\times& \left[|A_{k+}|^2-|A_{(k+1)-}|^2\right]\nn\\
&\times& F(-k,k+2,2,\sin^2{\Theta\over2})\label{eq:impart_helampl}\\
{\rm Im}T_{1-{1\over2},-1{1\over2}}&=&-\sin^3{\Theta\over2}8\pi q_\pi\sqrt{s}
\sum_{k\geq1}\frac{k^2(k+1)^2(k+2)^2}{12}\nn\\
&\times&\left[|B_{k+}|^2-|B_{(k+1)-}|^2\right]\nn\\
&\times& 
F(-k+1,k+3,4,\sin^2{\Theta\over2})\\
{\rm Im}T_{1{1\over2},-1{1\over2}}&=&
\sin^2{\Theta\over2}\cos{\Theta\over2}8\pi q_\pi\sqrt{s}
\sum_{k\geq1}\frac{k(k+1)^2(k+2)}{2}\nn\\
&\times&{\rm Re}\left[B_{k+}A^*_{k+}-B_{(k+1)-}A^*_{(k+1)-}\right]\nn\\
&\times& 
F(-k+1,k+3,3,\sin^2{\Theta\over2}),
\eeqn
where $q_\pi$ is the c.m. pion three-momentum and the hypergeometric function 
is defined as 
\beqn
F(a,b,c,x)\;=\;1+\frac{ab}{c}\frac{x}{1!}
+\frac{a(a+1)b(b+1)}{c(c+1)}\frac{x^2}{2!}+\dots
\eeqn

We keep only 
few first multipoles in this infinite series, namely 
$A_{0+}$, $A_{1+}$, $B_{1+}$, $A_{2-}$, $B_{2-}$ 
which obtain their leading contributions from threshold pion production, 
$\Delta(1232)$ and $D_{13}(1520)$ resonances. The result 
for $B_n$ reads:
\beqn
&&B_n=-8\pi m q_\pi \frac{st^2}{(s-M^2)^2}I_0
\frac{\sqrt{2\varepsilon(1-\varepsilon)}\sqrt{1+\tau}}
{\tau G_M^2+\varepsilon G_E^2}\nn\\
&&\cdot
\left\{
\left(\frac{E_{lab}}{M}F_2-F_1\right)\right.\nn\\
&&\;\;\;\cdot
\left[|A_{0+}|^2-|A_{1-}|^2+4(|A_{1+}|^2-|A_{2-}|^2)\right]
\nn\\
&&-\frac{6\sqrt{s}}{M}F_1\,{\rm Re}
\left(B_{1+}A_{1+}^*-B_{2-}A_{2-}^*\right)
\label{eq:bn_helmult}\\
&&\;-\frac{3}{2}\sin^2{\Theta\over2}
\left[
\left(\frac{E_{lab}}{M}F_2-F_1\right)\right.
\nn\\
&&\;\;\;\;\;\;\;\;\;\;\;\;\;\;\;\;\;\;\;\left.\cdot\left(
4(|A_{1+}|^2-|A_{2-}|^2)+|B_{1+}|^2-|B_{2-}|^2
\right)\right.\nn\\
&&
\;\;\;\;\;\;\;\;\;\;\;\;\;\;\;\;\;\;
+2\left(\frac{k}{M}F_2-\frac{E}{M}F_1\right)\nn\\
&&\left.\;\;\;\;\;\;\;\;\;\;\;\;\;\;\;\;\;\;\;\;\;\;
\cdot
{\rm Re}\left(B_{1+}A_{1+}^*-B_{2-}A_{2-}^*\right)
\right]
\Bigg\}.\nn
\eeqn

The helicity multipoles are related to the electromagnetic multipoles 
$E_{l+,(l+1)-}$ $M_{l+,(l+1)-}$ as 
\beqn
E_{0+}&=&A_{0+},\;\;\;\;\;\;\;\;\;\;\;\;M_{1-}\,=\,A_{1-},
\eeqn
and for $l\geq1$
\beqn
E_{l+}&=&\frac{1}{l+1}\left[A_{l+}+\frac{l}{2}B_{l+}\right]\nn\\
M_{l+}&=&\frac{1}{l+1}\left[A_{l+}-\frac{l+2}{2}B_{l+}\right]\nn\\
E_{(l+1)-}&=&-\frac{1}{l+1}\left[A_{(l+1)-}-\frac{l+2}{2}B_{(l+1)-}\right]\nn\\
M_{(l+1)-}&=&-\frac{1}{l+1}\left[A_{(l+1)-}+\frac{l}{2}B_{(l+1)-}\right]
\eeqn

It is informative to consider the contributions of $E_{0+}$ and $M_{1+}$ which 
are dominant at low energies. Neglecting other multipoles, we get:
\beqn
&&B_n=-8\pi m q_\pi \frac{st^2}{(s-M^2)^2}I_0
\frac{\sqrt{2\varepsilon(1-\varepsilon)}\sqrt{1+\tau}}
{\tau G_M^2+\varepsilon G_E^2}\nn\\
&&\cdot
\left\{
\left(\frac{E_{lab}}{M}F_2-F_1\right)
\left[|E_{0+}|^2+|M_{1+}|^2\right]\right.\nn\\
&&+
\left(
\frac{3\sqrt{s}}{M}F_1
+3\sin^2{\Theta\over2}\frac{(\sqrt{s}-M)^2}{2\sqrt{s}M}
\left[\frac{\sqrt{s}+M}{M}F_2-F_1\right]\right)\nn\\
&&\;\;\;\;\;\cdot
|M_{1+}|^2
\Bigg\}.
\eeqn

\begin{figure}[h]
{\includegraphics[height=8cm]{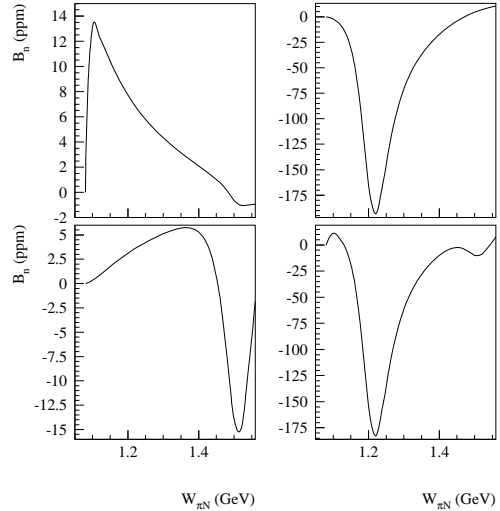}}
\caption{Beam normal spin asymmetry for the reaction $\vec{e}+p\to e+p$ 
in the QRCS approximation at fixed c.m. scattering angle 
$\Theta_{cm}=120^o$ as function of 
the $\pi N$ invariant mass $W_{\pi N}$. The contribution of the $E_{0+}$ 
multipole (upper left panel), $E_{1+}$ and $M_{1+}$ (upper right panel), 
$E_{2-}$ and $M_{2-}$ (lower left panel) are shown separately. The full result 
within the QRCS approximation is shown in the lower right panel.}
\label{fig:ssa_pin_energy}
\end{figure}

\begin{figure}[h]
{\includegraphics[height=8cm]{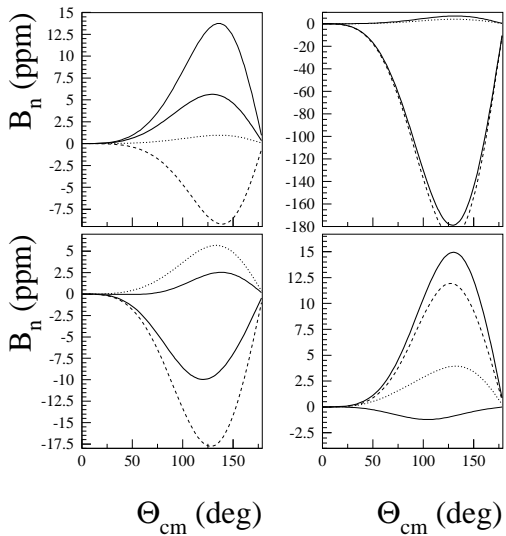}}
\caption{Beam normal spin asymmetry on the proton as function of c.m. 
scattering angle at 
different values of $lab$ energy: 200 MeV (upper left panel), 300 MeV (upper 
right panel), 570 MeV (lower left panel) and 855 MeV (lower right panel). 
The different lines represent $E_{0+}$ multipole contribution (thin solid 
lines), $E_{1+}$ and $M_{1+}$ contribution (dashed lines), $E_{2-}$ and 
$M_{2-}$ contribution (dotted lines), and the sum of all (thick solid lines)}
\label{fig:ssa_pin_angle}
\end{figure}

\begin{figure}[h]
{\includegraphics[height=8cm]{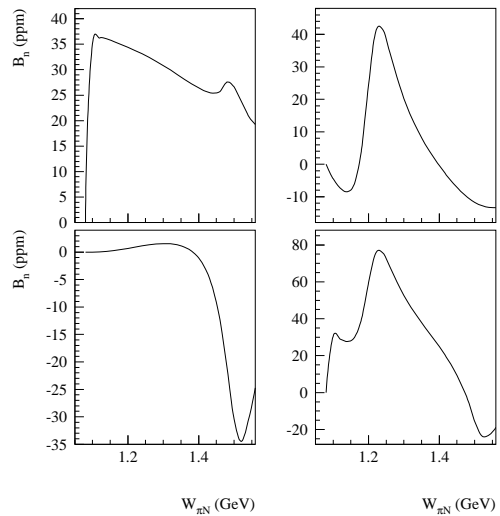}}
\caption{Beam normal spin asymmetry for the reaction $\vec{e}+n\to e+n$
in the QRCS approximation as function of the $\pi N$ invariant mass 
$W_{\pi N}$. Notation as in Fig. \ref{fig:ssa_pin_energy}}
\label{fig:ssa_pin_energy_neutron}
\end{figure}

The first term in the brakets is negative for 
$E_{lab}\leq{M\over\kappa}\approx0.45$ GeV for the proton target, and allways 
negative for the neutron target. Due to the overall minus sign, 
the beam normal spin asymmetry necessarily obtains a positive contribution 
from the threshold pion production. Moving towards the $\Delta(1232)$ resonance 
position, we see that the dominant contribution now comes from the second term. 
The factor multiplying the $|M_{1+}|^2$ term is allways positive for the 
proton and negative for the neutron. Therefore one expects that the $\Delta$ 
resonance give a large negative contribution to $B_n$ on the proton target, 
and a positive one on the neutron target. 
We use the MAID2003 multipoles as input to Eq.(\ref{eq:bn_helmult}) and present 
in Fig. \ref{fig:ssa_pin_energy} the energy dependence of the beam normal spin 
asymmetry for the proton target over the resonance region. 
This result can be compared to the results of 
Refs.\cite{marcbarbara, kobushkin}. If confronted to the full calculation 
\cite{marcbarbara}, we find agreement with their findings in the 
QRCS approximation. In Ref. \cite{kobushkin}, the authors do nominally the same 
approximation as we do, keeping the $\ln^2(Q^2/m^2)$ and the RCS part of the 
hadronic tensor only but arrive to the same sign of the contributions of the 
$E_{0+}$ and $M_{1+}$ multipoles for the proton. 
They are furthermore able to describe both 
low energy backward angle data from SAMPLE and intermediate energy and angle 
data from MAMI within the same approximation. It has been shown in 
\cite{marcbarbara} that the QRCS approximation does work well at backward 
angles (in the sense that it represents the dominant part of the full 
integration range) but drops short at forward angles, where the exchange of 
at least one hard virtual photon is important (for very forward angles, see 
\cite{afanas_qrcs}). Therefore, even if the QRCS approximation did work in the 
forward regime, one should not take this success too seriously, since the 
neglected double or single virtual Compton scattering effects are important. 
We show the angular distributions of $B_n$ on the proton target for 
different values of the electron $lab$ energy in Fig. \ref{fig:ssa_pin_angle}.
In Fig. \ref{fig:ssa_pin_energy_neutron}, we display the energy distributions
of $B_n$ on the neutron over the resonance region.

It is necessary to note that the resonance 
region does not seem to be favorable for the QRCS approximation. In this 
approximation, the value of the hadronic amplitude is taken at the largest 
energy available and is not integrated over the full spectrum. Therefore, if 
one goes above a resonance position, it is only tail of the resonance that 
contribute, which causes the asymmetry to drop quite fast (see the sharp 
resonance behaviour at $\Delta(1232),D_{13}(1520)$ in 
Fig.\ref{fig:ssa_pin_energy}). In a full 
calculation, also intermediate energies of the hadronic system do contribute 
and the asymmetry does not have such a sharp energy dependence. So, it is 
preferable to use the QRCS approximation in the region where the energy 
dependence is rather smooth. Furhtermore, the quality of the QRCS approximation 
is function of the scattering angle, as well. It works better at backward 
angles and worse at forward angles. The reason for this is clear: taking the 
limit of the soft intermediate electron, $k_1\approx0$ and neglecting the 
dependence of the 
hadronic and leptonic tensors on $k_1$ corresponds to performing Taylor 
expansion 
of the integration result in powers of $\ln{Q^2\over m^2}$ and neglecting all 
the terms beyond the first leading log term $\sim\ln^2{Q^2\over m^2}$. The 
larger $Q^2$, the better the approximation works. This 
behaviour was first observed in Ref.\cite{marcbarbara}. 
Instead, results of Ref.\cite{kobushkin} 
suggest that the loop integral in $B_n$ is completely dominated by the QRCS 
region in all the kinematics (backward regime for SAMPLE data and 
rather forward regime for MAMI data). 
\subsection{Regge regime: 2$\pi$ exchange}
The calculation of Ref. \cite{afanas_erratum} estimates $B_n$ as 
\beqn
B_n\sim\sigma_{tot}\ln(-t/m^2),
\eeqn
which is based on taking the RCS amplitude in the exact forward limit. If one 
allows the momentum transfer to be non-zero, one however obtains the 
contribution from photon helicity flip amplitudes, which is the result of 
this work. Though it is logical to expect that these amplitudes be suppressed 
at low values of $t$, there is a competing factor of $\sim\ln^2{Q^2\over m^2}$, 
which can be of the order of several hundreds. In this subsection, we provide 
an estimate of such a contribution at forward kinematics. 
The combinations of the RCS amplitudes $B_1+B_3$, $B_2+B_4$ 
in Regge regime are known to have the quantum numbers of a 
scalar isoscalar exchange in the $t$-channel which was successfully described 
by $\sigma$-meson \cite{lvov} or equivalently, by two pion exchange in the 
$t$-channel \cite{jarcs}. Since the effective $\sigma$-meson exchange does not 
result in a non-zero imaginary part in the $s$-channel, one should use the 
two pion exchange mechanism accompanied by multiparticle intermediate state in 
the $s$-channel. In this section we estimate $B_1+B_3$ and $B_2+B_4$ through 
$\pi N$ and $\rho N$ intermediate states
contributions within the ``minimal'' Regge model for $\pi$ and $\rho$ 
photoproduction \cite{marcmichel_piprod} where reggeized description of 
high energy pion 
production is obtained by adding the $t$-channel meson exchange amplitude and 
(in the case of $\gamma p$ reaction) $s$-channel Born amplitude which is 
necessary to ensure gauge invariance. The reggeization procedure naturally 
leads to the substitution of each $t$-channel Feynman propagator by its 
Regge counterpart,
${1\over t-m_\pi^2}\to {\cal P}_\pi^R(\alpha_\pi(t))$, with 
\beqn
{\cal{P}}_R^\pi\,=\,
\left(\frac{s}{s_0}\right)^{\alpha_\pi(t)}
\frac{\pi\alpha'_\pi}{\sin{\pi\alpha_\pi(t)}}
\frac{1}{\Gamma(1+\alpha_\pi(t))}\,,
\eeqn
with $\alpha_\pi(t)=\alpha'_\pi(t-m_\pi^2)$ and $\alpha'_\pi=0.7$ GeV$^{-2}$. 
Gauge invariance requires the $s$-channel piece to be reggeized in the 
same way, i.e., to be multiplied by  
$(t-m_\pi^2){\cal P}_\pi^R(\alpha_\pi(t))$. For compactness, we will use the 
shorthand \cite{lvov},
\beqn
A_1\equiv\frac{1}{t}\Big[B_1+B_2+\nu(B_2+B_4)\Big]
\eeqn
for the combination of the $B_i$'s which enters the result for the asymmetry.
Here, we list the results of the calculation of $A_1$:
\beqn
{\rm Im}A_1^{\pi N}&=&\frac{2m_\pi^2C_\pi^2}{M(s-M^2)}(B_0^\pi-M^2A_0^\pi),
\label{eq:a1_pi}\\
{\rm Im}A_1^{\rho N}&=&\frac{m_\rho^2C_\rho^2}{2M(s-M^2)}\label{eq:a1_rho}
\left\{{M^2\over s-M^2}C_0^\rho\right.\nn\\
&&+\,\left.{s-3M^2
+5m_\rho^2\over2}B_0^\rho+m_\rho^2{s+M^2\over2}A_0^\rho
\right\},\nn
\eeqn
where $C_\pi=2\sqrt{2}Me{f_{\pi NN}\over m_\pi}$ with $f_{\pi NN}^2/4\pi=0.08$,
and $C_\rho=2\sqrt{2}Me{f_{\pi NN}\over m_\pi}{f_{\rho\pi\gamma}\over m_\pi}$ 
with $f_{\rho\pi\gamma}=0.103$ \cite{marcmichel_piprod}. Where possible, 
$m_\pi$ and $Q^2$ were neglected as compared to $s$, $M$, $m_\rho$ in order to 
simplify the final expression. The scalar integrals 
$A_0$, $B_0$, $C_0$ for both $\pi N$ and $\rho N$ contributions are given in 
the Appendix. \\
\indent
The result of Eq.(\ref{eq:a1_pi}) leads to negligibly small contribution to the 
$B_n$, of the order of $10^{-2}$ ppm for energies between 6 and 45 GeV, 
which is to be compared to $\approx5$ ppm from Ref. \cite{afanas_erratum}. 
There are three suppression factors which are 
responsible for this small result. Firstly, the Reggeization procedure which 
leads to suppression in energy $\nu^{\alpha_\pi(t)}$. Secondly, 
the amplitude $A_1$ is defined as the combination 
${1\over -t}[B_1+B_3+\nu(B_2+B_4)]$, and the singularity ${1\over t}$ in 
the definition only cancels if taking the gauge invariant combination as 
described in \cite{marcmichel_piprod}. Therefore, we obtain an additional 
suppression factor of $t$. The other feature of the result of 
Eqs.(\ref{eq:a1_pi},\ref{eq:a1_rho}) is that both contributions are 
suppressed by factors ${m_\pi^2\over s-M^2}$ and $m_\rho^2\over s-M^2$, 
respectivelyly. In the case of the pion, it is interesting to observe this 
fact in view of somewhat surprising success of the effective field description 
of the SAMPLE data point on $B_n$ without pion contribution. This might give 
a hint that the pion contribution to $B_n$ at low energies is suppressed by 
the pion mass, if calculated to the same order.

\subsection{Hard regime: GPD's}
Recently, beam normal spin asymmetry was studied in the hard scattering regime 
\cite{jamarcguichon} in the framework of the handbag mechanism. In this 
picture, the process factorizes into the hard part (electron scattering off 
an on-shell quark with the exchange of two spacelike photons) and the soft 
part which is 
parametrized by the generalized parton distributions (GPD's). However, 
one cannot access the QRCS kinematics here, since it is impossible 
for an on-shell quark to absorb (emit) a hard real photon and remain on-shell. 
As it has been noticed before, the QRCS kinematics selects the RCS amplitudes 
with the photon helicity flip, i.e. the helicity changes by two units. 
In the context of the GPD's, helicity amplitudes with the helicity changed by 2 
units were considered in several works. In the collinear kinematics, 
this requires that scattering occurs on a parton of spin 1, that is gluon. 
The corresponding gluon helicity flip GPD's were introduced in \cite{gluonGPD} 
\beqn
&&\frac{1}{x}\int\frac{d\lambda}{2\pi}e^{i\lambda x}
< p'S'\,|\,F^{\left(\mu\alpha\right.}(-\frac{\lambda}{2}n)
F^{\left.\nu\beta\right)}(-\frac{\lambda}{2}n)\,|\,pS >\nn\\
&&=H_{T_g}(x,\xi)
\bar{N}'(p',S')
\frac{P^{\left(\right.\left[\mu\right.}q^{\left.\alpha\right]}i\sigma^{\left.\nu\beta\right)}}{M}
N(p,S)
\nn\\
&&+E_{T_g}(x,\xi)
\bar{N}'(p',S')
\frac{P^{\left(\right.\left[\mu\right.}q^{\left.\alpha\right]} 
\gamma^{\left[\nu\right.}q^{\left.\beta\right]\left.\right)}}{M^2}N(p,S)
\eeqn
where $H_{T_g}(x,\xi), E_{T_g}(x,\xi)$ are the double helicity flip gluon 
GPD's, $q=p'-p$, and the matrix element is taken between the initial and final 
nucleon states with the momenta $p$ and $p'$, and spins $S$ and $S'$, 
respectively. The integration is performed along the light-cone vector 
$n^\mu=(1,0,0,1)$. $x$ and $\xi$ are the longitudinal parton momentum fraction 
and skewness parameter, respectively. 
Furthermore, $[\mu\alpha]$ and $[\nu\beta]$ are antisymmetric 
pairs of indices, while $(\dots)$ means symmetrization and removal of the 
trace. These operations are essential since the product of operators must 
transform as irreducible representations of the Lorentz group.
This requirement rules out the possibility to see 
these GPD's in the QRCS kinematics, since it is exactly the trace of the RCS 
tensor that contributes to $B_n$ in the QRCS regime, c.f. 
Eq.(\ref{eq:tensor_contraction}). So, our conclusion is that in the QRCS 
kinematics, contributions from generalized parton distributions are ruled out. 

\section{Summary}
\label{sec:summary}
In summary, we presented a calculation of the beam normal spin asymmetry. 
This observable 
obtains its leading contribution from the imaginary part of the 
two photon exchange graph times the Born amplitude and is 
directly related to the imaginary part of doubly virtual Compton scattering
The resulting loop integral obtains a large contribution from the kinematics 
when both exchanged photons are nearly real and collinear to the external 
electrons. We adopt the QRCS or equivalent photons approximation which allows 
to take the hadronic part out of the integral and 
to perform the integration over the electron phase space analytically. 
For the hadronic part, we use the full real Compton scattering amplitude 
and show, that only photon helicity flipping amplitudes contribute in this 
observable in the QRCS approximation. At low energies, we relate this helicity 
flipping Compton amplitude to the $\pi N$ multipoles and discuss the 
contributions of different multipoles. 

At high energies and forward scattering angles (Regge regime), we provide 
an explicite calculation which is due to two pion exchange in the $t$-channel. 
The resulting values of the asymmetry are of the order $10^{-2}$ ppm for the 
energies in the range $6-45$ GeV. We conclude that the double 
logarithmic enhancement does not dominate $B_n$ in forward regime since it 
comes with helicity-flip Compton amplitude which highly suppresses this 
behaviour. 

Finally, we consider possible contributions to $B_n$ in the hard 
kinematics, among which 
contributions from quark and gluon GPD's and show that such contributions are 
ruled out in the QRCS kinematics. 
\begin{acknowledgments} The author is grateful to Prof. M.M.Giannini for 
continuous support and to Dr. M.J. Ramsey-Musolf and D. O'Connell for 
numerous discussions and for thorough reading of the manuscript. 
The work was supported by Italian MIUR and INFN, and by US Department of 
Energy Contract DE-FG02-05ER41361.
\end{acknowledgments}
\section{Appendix A: Integrals over electron phase space}
In this section we present calculation of the integrals over electron phase 
space appearing in the QRCS approximation. First we calculate the scalar 
integral $I_0$,
\beqn
I_0\;=\;\int_0^{k_{thr}}\frac{k_1^2dk_1}{2E_1(2\pi^3)}
\int\frac{d\Omega_{k_1}}{(k-k_1)^2(k'-k_1)^2},
\eeqn
where the upper integration limit $k_{thr}$ corresponds to the inelastic 
threshold (i.e. pion production), 
$k_{thr}=\sqrt{\frac{(s-(M+m_pi)^2)^2}{4s}-m^2}$. We 
next introduce integration over the Feynman parameter 
${1\over ab}=\int_0^1{dx\over [a+(b-a)x]^2}$. We chose the polar axis such as 
$\vec{k}_1\cdot(\vec{k}-x\vec{q})=k_1|\vec{k}-x\vec{q}|\cos\Theta_1$ with 
$|\vec{k}-x\vec{q}|^2=k^2+x(1-x)t$ and perform angular integration.
We furthermore change integration over $dk_1$ 
to integration over dimensionsless $z=E_1/E$
\beqn
I_0&=&
\frac{1}{-8\pi^2t}\int_{m\over E}^{{E_{thr}}\over E}
\frac{dz}{\sqrt{z^2-\frac{m^2}{E^2}-\frac{4m^2}{t}(1-z)^2}}\\
&&\;\;\;\;\;\;\cdot
\ln\frac{\sqrt{z^2-\frac{m^2}{E^2}-\frac{4m^2}{t}(1-z)^2}+\sqrt{z^2-\frac{m^2}{E^2}}}
{\sqrt{z^2-\frac{m^2}{E^2}-\frac{4m^2}{t}(1-z)^2}-\sqrt{z^2-\frac{m^2}{E^2}}},
\nn
\eeqn
\indent
To perform the integration 
over the electron energy, we follow here the main details of the calculation 
in Appendix A of Ref.\cite{afanas_qrcs}. The result reads
\beqn
I_0\;=\;\frac{1}{-32\pi^2t}
\left\{
\ln^2\left(\frac{-t}{m^2}\frac{E_{thr}^2}{E^2}\right)\,+\,
8Sp\left({{E_{thr}}\over E}\right)\right\},
\eeqn
where $Sp(x)$ is the Spence or dilog function, 
$Sp(x)=-\int_0^1{dt\over t}\ln(1-xt)$ with $Sp(1)=\pi^2/6$. In the high 
energy limit, ${{E_{thr}}\over E}\to1$, we recover the result of 
Ref.\cite{afanas_qrcs}.

\section{Appendix B: Scalar integrals for helicity flip amplitude}
The vector and tensor integrals can be reduced 
to the scalar ones by means of standard methods \cite{passarino_veltman}. 
The remaining integrals to be calculated are the two, three, and four-point 
scalar integrals. Here we are only interested in the imaginary part of these, 
therefore there are only three integrals with non-zero imaginary part:
the two-point integral
\beqn
C^\pi_0&=&{\rm Im}\int\frac{d^4p_\pi}{(2\pi)^4}
\frac{1}{p_\pi^2-m_\pi^2}\frac{1}{(P+K-p_\pi)^2-M^2}\nn\\
&=&\frac{1}{8\pi}\frac{|\vec{p}_\pi|}{\sqrt{s}},
\eeqn
with $|\vec{p}_\pi|=\sqrt{\frac{(s-M^2+m_\pi^2)^2}{4s}-m_\pi^2}$;
the three-point one
\beqn
B^\pi_0&=&{\rm Im}\int\frac{d^4p_\pi}{(2\pi)^4}
\frac{1}{p_\pi^2-m_\pi^2}\frac{1}{(k-p_\pi)^2-m_\pi^2}\nn\\
&&\;\;\;\;\;\;\;\;\cdot
\frac{1}{(P+K-p_\pi)^2-M^2}\nn\\
&=&-\frac{1}{8\pi(s-M^2)}\ln\frac{2E_\pi}{m_\pi},
\eeqn
and finally, the four-point integral:
\beqn
A^\pi_0&=&{\rm Im}\int\frac{d^4p_\pi}{(2\pi)^4}
\frac{1}{(k-p_\pi)^2-m_\pi^2}\frac{1}{p_\pi^2-m_\pi^2}\nn\\
&&\;\;\;\;\;\;\;\;\cdot
\frac{1}{(k'-p_\pi)^2-m_\pi^2}\frac{1}{(P+K-p_\pi)^2-M^2}\nn\\
&=&\frac{1}{8\pi Q^2(s-M^2+m_\pi^2)}
\nn\\
&\cdot&
\frac{1}{\sqrt{1+{{4m_\pi^2E^2}\over Q^2p_\pi^2}}}
\ln\frac{\sqrt{1+{{4m_\pi^2E^2}\over Q^2p_\pi^2}}+1}
{\sqrt{1+{{4m_\pi^2E^2}\over Q^2p_\pi^2}}-1}.
\eeqn
\indent
These integrals should however be reggeised as described in Section 
\ref{sec:results} by substituting the Regge propagator instead of the Feynman 
one. Denoting 
$t_1=(k-p_\pi)^2$ and $t_2=(k'-p_\pi)^2$ the momentum transferred by the pions 
in the $t$-channel, we have for the reggeized version of scalar integrals:
\beqn
(C^\pi_0)^R&=&\frac{1}{32\pi^2}\frac{p_\pi}{\sqrt{s}}\int d\Omega_\pi
(t_1-m_\pi^2){\cal P}_\pi^R(\alpha_\pi(t_1))\nn\\
&&\;\;\;\;\;\;\;\;\;\;\;\;\;\;\;\;\;\;
\cdot(t_2-m_\pi^2){\cal P}_\pi^R(\alpha_\pi(t_2))
\eeqn
\beqn
(B^\pi_0)^R&=&\frac{1}{32\pi^2}\frac{p_\pi}{\sqrt{s}}\int d\Omega_\pi
(t_1-m_\pi^2)\\
&&\;\;\;\;\;\;\;\;\;\;\;\;\;\;{\cal P}_\pi^R(\alpha_\pi(t_1))
{\cal P}_\pi^R(\alpha_\pi(t_2)),\nn
\eeqn
\beqn
\!\!\!\!\!\!(A^\pi_0)^R
\,=\,\frac{1}{32\pi^2}\frac{p_\pi}{\sqrt{s}}\!\!\int \!\!d\Omega_\pi
{\cal P}_\pi^R(\alpha_\pi(t_1))
{\cal P}_\pi^R(\alpha_\pi(t_2))
\eeqn
\indent
Similarly, in the case of the $\rho$-exchange in the $s$-channel, the integrals
 with non-zero imaginary part are:
\beqn
C^\rho_0&=&{\rm Im}\int\frac{d^4p_\rho}{(2\pi)^4}
\frac{1}{p_\rho^2-m_\rho^2}\frac{1}{(P+K-p_\rho)^2-M^2}\nn\\
&=&\frac{1}{8\pi}\frac{|\vec{p}_\rho|}{\sqrt{s}},\nn
\eeqn
\beqn
B^\rho_0&=&{\rm Im}\int\frac{d^4p_\rho}{(2\pi)^4}
\frac{1}{p_\rho^2-m_\rho^2}\frac{1}{(k-p_\rho)^2-m_\pi^2}\nn\\
&&\;\;\;\;\;\;\;\;\;\;\;\;\;\;\cdot\frac{1}{(P+K-p_\rho)^2-M^2}\nn\\
&=&-\frac{1}{8\pi(s-M^2)}\ln\frac{2E_\rho(s-M^2)}{Mm_\rho^2},\nn
\eeqn
\beqn
A^\rho_0&=&{\rm Im}\int\frac{d^4p_\rho}{(2\pi)^4}
\frac{1}{(k-p_\rho)^2-m_\pi^2}\frac{1}{p_\rho^2-m_\rho^2}\nn\\
&&\;\;\;\;\;\;\;\;\;\;\;\;\;\;\cdot
\frac{1}{(k'-p_\rho)^2-m_\pi^2}\frac{1}{(P+K-p_\rho)^2-M^2}\nn\\
&=&\frac{1}{8\pi Q^2(s-M^2+m_\rho^2)}
\frac{1}{\sqrt{1+{{4\sigma^2}\over Q^2p_\rho^2}}}\nn\\
&\cdot&
\ln\frac{\sqrt{1+{{4\sigma^2}\over Q^2p_\rho^2}}+1}
{\sqrt{1+{{4\sigma^2}\over Q^2p_\rho^2}}-1},
\eeqn
with $|\vec{p}_\rho|=\sqrt{\frac{(s-M^2+m_\rho^2)^2}{4s}-m_\rho^2}$, and 
$\sigma^2=E^2m_\rho^2-EE_\rho(m_\rho^2-m_\pi^2)+{1\over4}(m_\rho^2-m_\pi^2)^2$.
If neglecting the pion mass, $\sigma={{Mm_\rho^2}\over 2s}$.
The reggeization procedure is the same as for $\pi N$- intermediate state.

\end{document}